# A REVIEW ON MEMRISTOR APPLICATIONS

Roberto Marani, Gennaro Gelao and Anna Gina Perri

Electronic Devices Laboratory, Electrical and Information Engineering Department,
Polytechnic University of Bari, via E. Orabona 4, Bari – Italy
annagina.perri@poliba.it

## ABSTRACT

*This article presents a review on the main applications of the fourth fundamental circuit element, named "memristor", which had been proposed for the first time by Leon Chua and has recently been developed by a team at HP Laboratories led by Stanley Williams. In particular, after a brief analysis of memristor theory with a description of the first memristor, manufactured at HP Laboratories, we present its main applications in the circuit design and computer technology, together with future developments.*

## 1. INTRODUCTION

Memristor, a concatenation of "*memory resistor*", is a type of passive circuit element that maintains a relationship between the time integrals of current and voltage across a two terminal element. In fact its resistance depends on the charge that flowed through the circuit. When current flows in one direction the resistance increases, in contrast when the current flows in opposite direction the resistance decreases. However resistance cannot go below zero. When the current is stopped, the resistance remains in the value that it had earlier.
It means that memristor "remembers" the current that last flowed through it.
In this paper we analyze the memristor theory. In particular a description of the first memristor, manufactured at HP Laboratories, is presented with its implementation in P-SPICE simulator. Finally we highlight its main applications in the circuit design and computer technology.
The presentation is organized as follows. Section 2 gives a brief description of memristor theory. The HP Memristor and its implementation in P-SPICE simulator is presented in Section 3, while the description of the main current applications of memristor is given in section 4. The conclusions and future developments are described in Section 5.

## 2. MEMRISTOR THEORY

Passive circuit theory can be thought of as a set of relationships between electromagnetic quantities:
1) Voltage *v*, defined as the change magnetic flux $\Phi$ with respect to time *t*;
2) Current *i*, defined as the change in electric charge *q* with respect to time;
3) Resistor *R*, defined as a linear relationship between voltage and current ($dv = Rdi$);
4) Capacitor *C*, defined as a linear relationship between voltage and electric charge ($dq = Cdv$);
5) Inductor *L*, defined as a linear relationship between magnetic flux $\Phi$ and current *i* ($d\Phi = Ldi$).
Of the six possible relationships, the only two electromagnetic quantities for which there are no pairings are magnetic flux and electric charge. However, in 1971 Leon Chua [1] hypothesized that mathematically, a fourth fundamental passive circuit element could exist, proposing the fourth component, called **memristor**, that binds the charge *q* to the linkage flux $\Phi$ (see Fig. 1, in which the electrical symbol of memristor is also indicated):

$$d\Phi = Mdq \qquad (1)$$

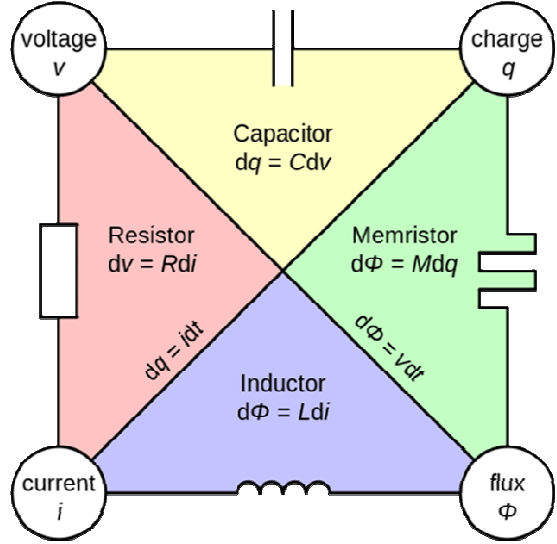

Figure 1. Memristor relationship (figure adapted from *Nature*).

In particular the memristor has originally been defined in terms of a non-linear functional relationship between the so-called *flux linkage φ(t)* and the amount of electric charge that has flowed through the device, *q(t)*:

$$f(\varphi(t),q(t)) = 0 \tag{2}$$

where *φ(t)* and *q* are time-domain integrals of memristor electric voltage *v* and electric current *i*, respectively:

$$\varphi(t) = \int_{-\infty}^{t} v(\tau)d\tau \tag{3}$$

and

$$q(t) = \int_{-\infty}^{t} i(\tau)d\tau \tag{4}$$

The variable flux linkage *φ(t)* is generalized from the circuit characteristic of an inductor, and does not represent a magnetic field here.
In Eq. (2) the derivative of one respect to the other depends on the value of one or the other.
In this way a memristor is characterized by its **memristance function**, which describes the charge-dependent rate of change of flux with charge:

$$M(q) = \frac{d\varphi}{dq} \tag{5}$$

Introducing Eq, (3) in Eq, (5), we have:

$$M(q(t)) = \frac{\frac{d\varphi}{dt}}{\frac{dq}{dt}} = R_M(q) \tag{6}$$

where $R_M(q)$ is the small-signal memristance defined in the operating point. In this way we have a charge-controlled or current-controlled memristor.
Analogously, introducing Eq, (4) in (5), we have:

$$M(\varphi(t)) = \frac{\frac{dq}{dt}}{\frac{d\varphi}{dt}} = G_M(\varphi) \qquad (7)$$

obtaining, in this way, a flux-controlled or voltage-controlled memristor.
The current-controlled memristor can be modeled via a classical resistor whose resistance is controlled by the time-domain integral of the current flowing through the memristor. Analogously, the voltage-controlled memristor behaves as a conductor whose conductance depends on the time-domain integral of terminal voltage.
In both cases, we need an electronically controlled resistor/conductor and an integrator.
Table 1 covers all meaningful ratios of differentials of $i$, $q$, $\varphi$ and $v$, while Fig. 2 shows the current-voltage characteristics for memristor, where you can easily notice its fundamental identifier, i.e. its pinched hysteresis loop.

Table 1

| Device | Units | Differential Equation |
|---|---|---|
| Resistor | Ohm | R = dv/di |
| Capacitor | Farad | C = dq/dv |
| Inductor | Henry | L = dφ/di |
| Memristor | Ohm | M = dφ/dq |

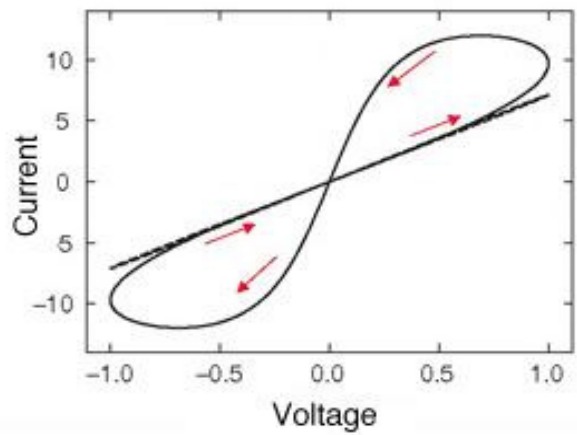

Figure 2. Current-voltage characteristics of a memristor.

With reference to Eq, (6), which can be written also in the following way;

v(t)= M(q(t)) i(t)     (8)

if $i(t) = 0$, we find $v(t) = 0$ and therefore $M(t)$ is constant (**memory effect**).
The power consumption characteristic recalls that of a resistor:

P(t) = V(t) I(t) = I²(t) M(q(t))     (9)

As long as $M(q(t))$ varies little, such as under alternating current, the memristor will appear as a constant resistor.
When $M(q(t))$ increases rapidly, however, current and power consumption will quickly stop.

Moreover Chua has hypothesized that this new device could have several applications in digital memory, analog electronics, neural networks and in [1] he proposed a possible realization of memristor using passive and active circuit elements, as shown in Fig. 3.

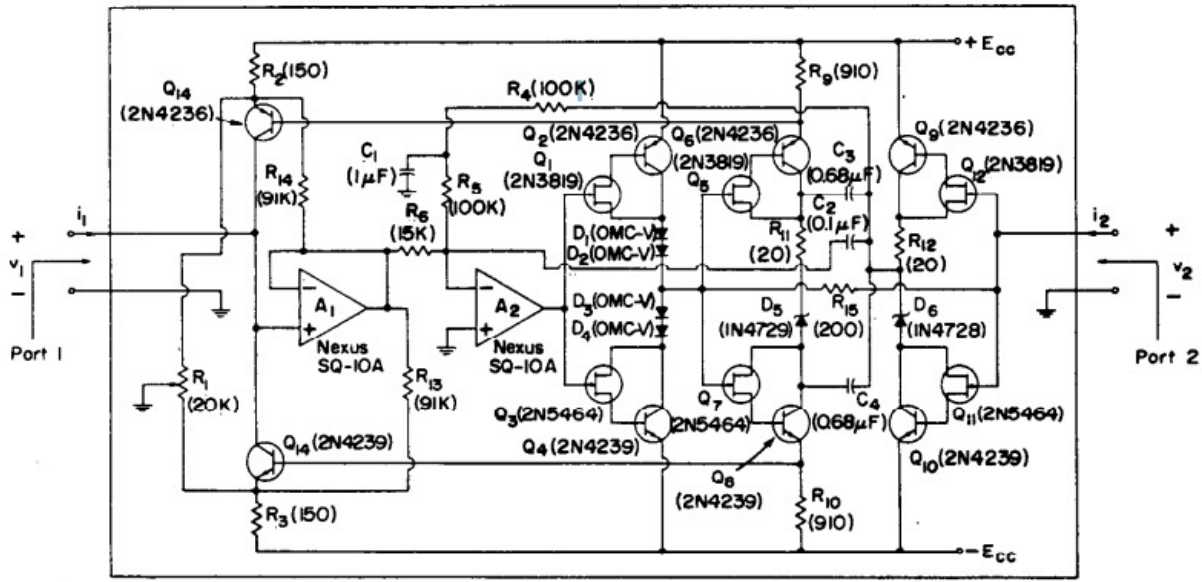

Figure 3. The possible realization of memristor proposed by Chua (*from [1]*).

While the three circuit elements (*R*, *C* and *L*) can be easily and reliably constructed, up to 2008 nobody has proposed a material or physical structure that shows the characteristics of a memristor.

## 3. HP MEMRISTOR AND ITS IMPLEMENTATION IN P-SPICE

In 2008, thirty-seven years after Chua theory [1], at HP Labs the first memristor device has been realized [2]. The HP Memristor is a simple device, constructed of a $TiO_2$ thin film sandwiched between platinum electrodes, and relies on two physical characteristics of titanium dioxide.
The first characteristic is the sensitivity of $TiO_2$ conductivity to oxygen depletion. $TiO_2$ is normally an insulator but behaves as an n-type semiconductor when oxygen vacancies are introduced, forming $TiO_{2-x}$.
The second characteristic that the HP Memristor relies on is that of anionic migration, the tendency for oxygen vacancies within $TiO_2$ to drift with an applied electric field.
The physical structure of HP memristor, as already described, consists of a two-layer thin film of $TiO_2$, sandwiched between platinum contacts.
One of the layers is doped with oxygen vacancies and thus it behaves as a semiconductor. The second, undoped region, has an insulating property, as shown in Fig. 4, where *D* represents the device length (≈ 10 - 30 nm) and *w* the doped region length.

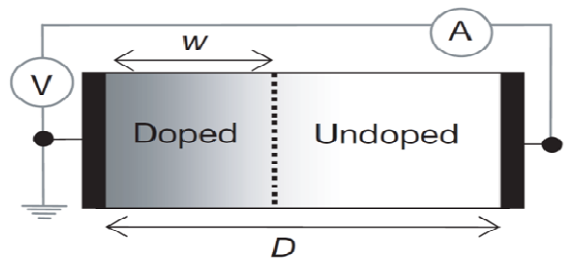

Figure 4. HP memristor (*from [3]*).

Therefore the resistance of the device, when w = D, denoted by $R_{ON}$, is low, while, for w = 0 the resistance, designated as $R_{OFF}$ (Fig. 5), is high.

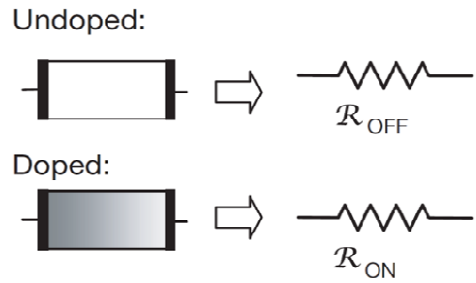

Figure 5. Resistance Naming Convention.

With an electric current passes in a given direction, the boundary between the two regions is moving in the same direction. The total resistance of the memristor is a sum of the resistances of the doped and undoped regions, whose values are dependant on the value of w, as shown in Fig. 6.

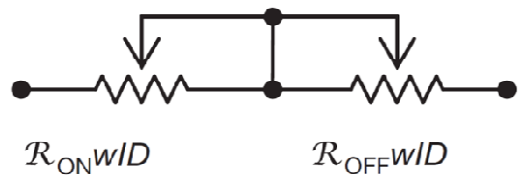

Figure 6. Effective Electrical Structure of the HP Memristor (*from [3]*)

Disconnecting the memristor from external voltage, the current stops flowing, the boundary stops moving, and the element remembers its resistance for a theoretically arbitrarily long time.
With reference to Fig. 6, the effective *i-v* behaviour of the structure can be represented as the following equation [4]:

$$v(t) = \left[ R_{ON} \frac{w(t)}{D} + R_{OFF}\left(1 - \frac{w(t)}{D}\right) \right]$$

and the memristance function, for $R_{ON} \ll R_{OFF}$, is expressed by [4]:

$$M(q(t)) = R_{OFF}\left(1 - \frac{\mu_v R_{ON}}{D^2} q(t)\right)$$

where $\mu_v$ is the mobility of dopants in the thin film.
In [5] it has been proposed an analogue model (Fig. 7) to be implemented in P-SPICE.

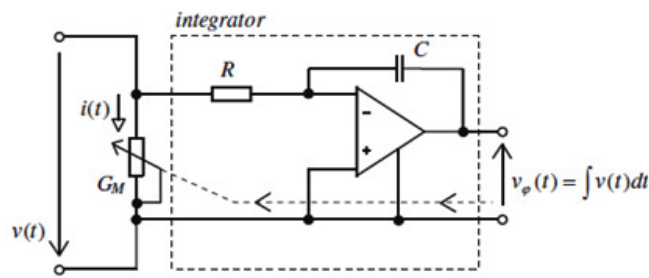

Figure 7. Memristor model implemented in P-SPICE (*from [5]*).

The model is based on a voltage-controlled memristor. It consists of an integrator producing a voltage $v_\varphi$ proportional to the flux $\varphi$ and a voltage-controlled conductance $G_M$, connected in parallel to the input terminals. The input current then is equal to the product of input voltage and conductance $G_M$ according to Eq. (7).

The integrator is realized classically using an OPAMP with a resistor $R$ connected to the inverting input and a capacitor $C$ in the feedback path. The integrator gain and the usable frequency band of the model are determined by the value of time constant $\tau = RC$.

This model may help to study and seize the behaviour characteristics of memristor, as several test cases and simulation results have shown in [5].

## 4. MAIN APPLICATIONS OF MEMRISTOR

The memristor has the potential to augment or enhance several areas of integrated circuit design and computing. Extensive literature regarding applications of the memristor has been produced since the HP announcement but few developments are particularly noteworthy.

One highly pervasive area where memristors may be applied is that of **non-volatile random access memory**, or **NVRAM** [6]. The memristor is seen as having significant potential in this area as the device exhibits memory does not require continuous power draw and consumes little physical area.

For **digital memory applications**, one bit of information can be stored using a single memristor. This can be achieved forcing the memristor to its extreme resistance values ($R_{ON}$ and $R_{OFF}$), each state corresponding to either a *1* or a *0*. DC voltages are used to set the resistance of a memristor element [6]. In order to read stored data, AC signals are utilized so that the stored data is not disturbed [6]. A crossbar arrangement is often used for this memory architecture. The crossbar topology is composed of a grid of horizontal and vertical traces [6]. At each intersection, a vertical trace is connected to a horizontal trace by a memristor. Thus any memristor can be programmed or read by applying a voltage to the necessary horizontal and vertical traces while letting the remaining traces float.

Memristor can be used as **associative memories** [7]. These memories map an input pattern to an output one according to the similarities of the input pattern to the pattern stored in the memory.

The fundamental cell is shown in Fig. 8, in which two memristors are programmed with complementary values: for the logic 0 state memX is set to high resistance while the memY is low resistance and viceversa for the logic 1 (cfr. Fig. 9).

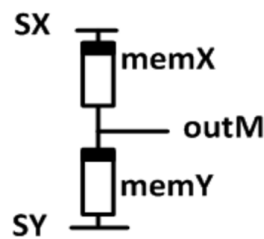

Figure 8. Fundamental cell of associative memories using memristors (*from [7]*).

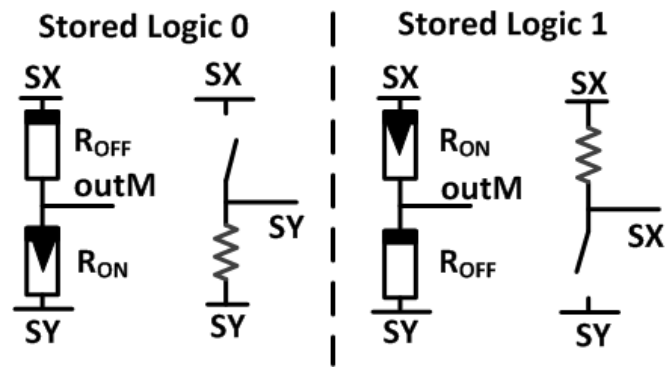

Figure 9. States of the fundamental cell (*from [7]*).

The state of the cell can be compared to a logic input 1 setting SY to high voltage and SX to low voltage, and viceversa for input 0. The matching is presented complemented as low voltage on output outM, as shown in Fig. 10.

|  | memX | memY | SX | SY | outM |
|---|---|---|---|---|---|
| Logic 0 | 0 | 1 | Searching Logic 1 | | 1 (Not Match) |
| | | | 0(GND) | 1(VDD) | |
| Logic 0 | 0 | 1 | Searching Logic 0 | | 0(Match) |
| | | | 1(VDD) | 0(GND) | |
| Logic 1 | 1 | 0 | Searching Logic 1 | | 0(Match) |
| | | | 0(GND) | 1(VDD) | |
| Logic 1 | 1 | 0 | Searching Logic 0 | | 1 (Not Match) |
| | | | 1(VDD) | 0(GND) | |

Figure 10. Matching logic of the associative cell (*from [7]*).

An associative array of 3x4 bit is illustrated in Fig. 11, in which the outputs are collected by the ML lines and sent to the line amplifiers. Since in this configuration each matching bit add current on the ML line, the total current is converted in a voltage output proportional to the matching bits.

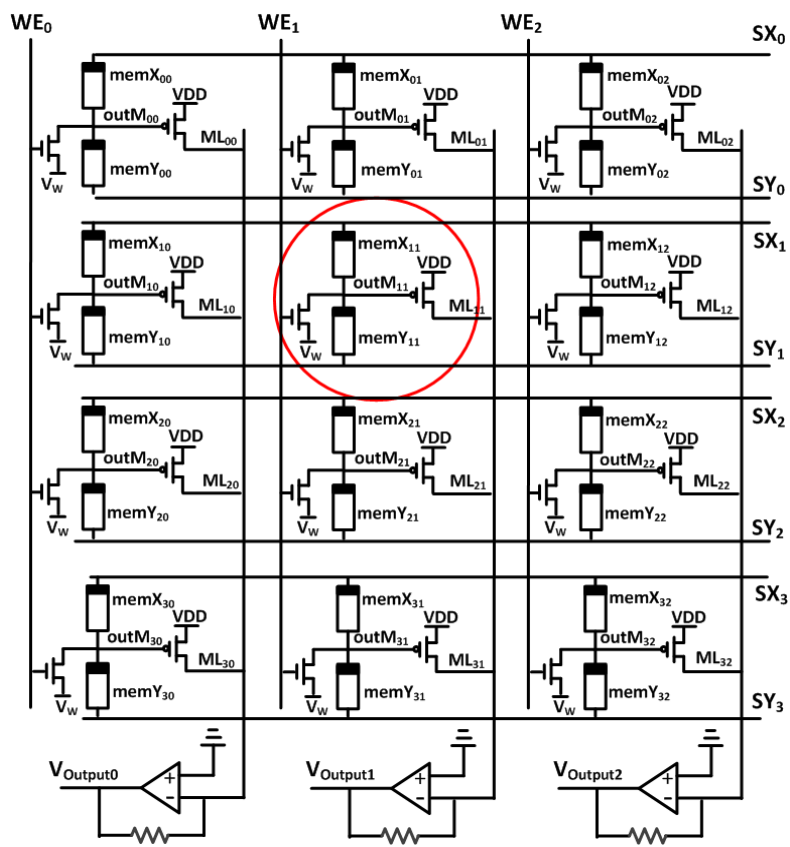

Figure 11. An associative memory system of 3x4 bits (*from [7]*).

Moreover memristor can possibly allow for **nano-scale low power memory** and **distributed state storage**, as a further extension of NVRAM capabilities.
While memristor can be used at its extreme resistance values in order to provide digital memory, it can also be made to behave in an analog manner.
One potential application of this behaviour is that of a dynamically adjustable **electric load** [8].

Thus, existing electronic circuit topologies with characteristics that depend on a resistance can be made with memristors that behave as variable **programmable resistances** [8].
This design approach has the potential to create electronically adjustable filters and amplifiers as shown in Fig. 12.

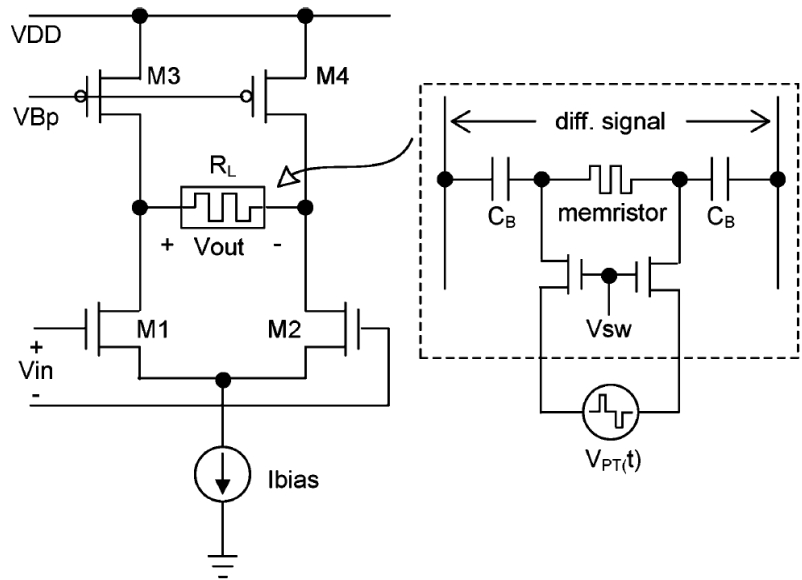

Figure 12. Programmable Gain Differential Amplifier (*from [8]*).

In particular, for analogic filter applications [9], the memristor is programmed with low frequency pulse to set its resistance using trick to separate high frequency signal path from the programming pulses, alike to those tricks used with varicap to separate high frequency signal path from bias voltage. In this way the circuit could be reconfigured while running.

In this applications it is very important that memristors are sensitive to low frequency pulses while they do no change their state under high frequency voltage, which has zero average value, since this allows the separation of the two operation: the programming of the memristor resistance values and the working as resistor.

Figs. 13 and 14 show a simulation of a second order filter where the Q-filter is controlled setting the memristor resistance [9].

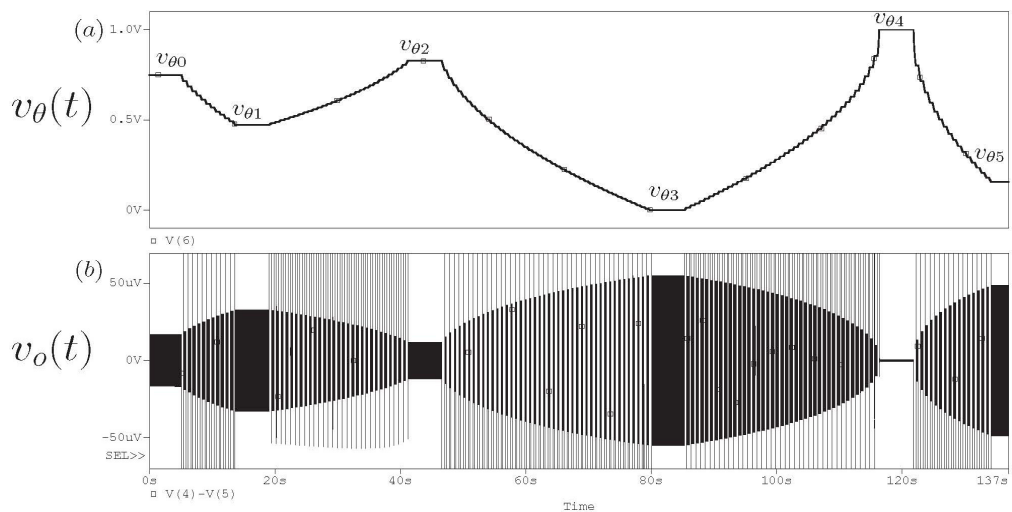

Figure 13 Memristor state variable (a) and filter output voltage (b) (*from [9]*).

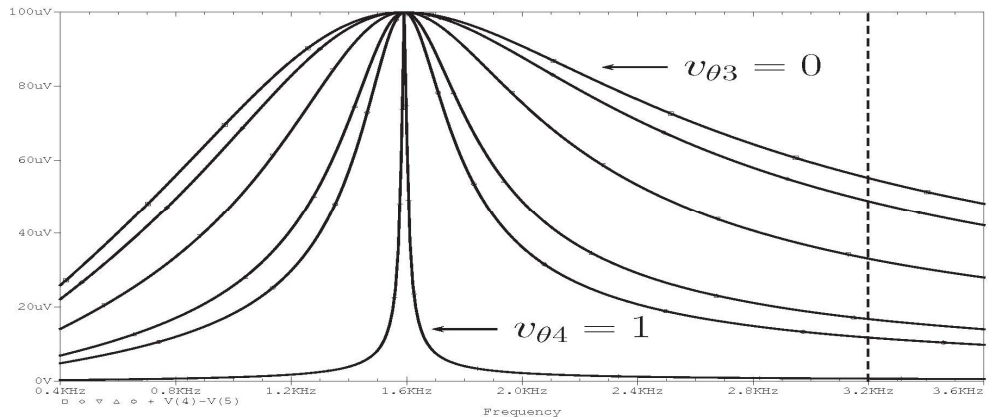

Figure 14. Filter transfer function in log scale as the controlling voltage $v_\theta$ changes during the simulation. Only the curves corresponding to the extreme voltages 0 V e 1 V are labelled (*from [9]*).

A common operation involved in image recognition and other image processing techniques is that of **edge detection**.

**Edge detection** identifies where large changes in a digital image occur, such as the outline of an object. However, edge detection is notoriously computationally intensive.

One application of memristors identified by [6] is that of performing edge detection using a memristance grid. An example of this operation can be seen below in Fig. 15.

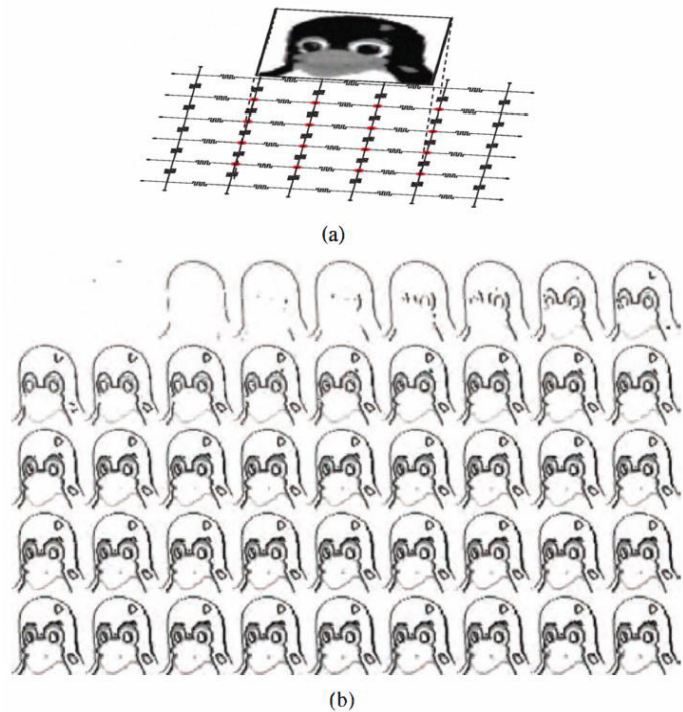

Figure 15. Memristor Based Edge Detection (*from [6]*).

The light intensity of the original photo is applied as voltages to points on a grid of Memristors, as shown in Fig. 15a. The resulting changes in resistance across the grid over time are then shown in the frames in Fig. 15b. Thus, after the system has been allowed to settle, the edge detected image can be recovered by measuring the resistance of each element in the grid [6].

A recent application is Memristor as **switched on Radio-Frequency (RF) antenna** [10]. In fact memristors can be used in reconfigurable RF antenna when they are used as switches between conductive antenna elements. Separation between programming mode and resistive mode is again obtained by the large difference in frequency of the two signals.

Figs. 16 and 17 show respectively the geometry and design results of a band-switching reconfigurable antenna designed for operation at 2.308 GHz and 3.143 GHz using memristors as switches between copper elements of the antenna. Joining or splitting electrically the elements of the antenna a change in the resonance frequency is obtained in this case, but more sophisticated examples could be obtained in phased array antenna where also the angular pattern could be controlled.

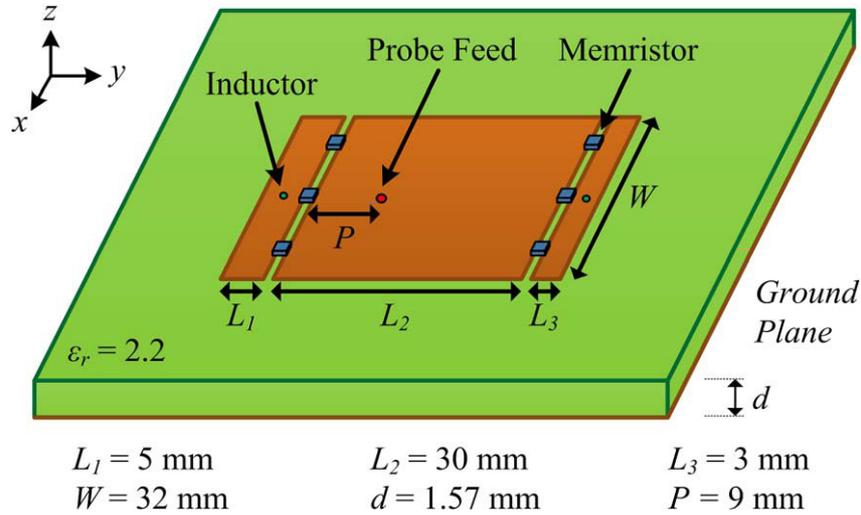

$L_1$ = 5 mm  $L_2$ = 30 mm  $L_3$ = 3 mm
$W$ = 32 mm  $d$ = 1.57 mm  $P$ = 9 mm

Figure 16. Geometry of the reconfigurable antenna (*from [10]*).

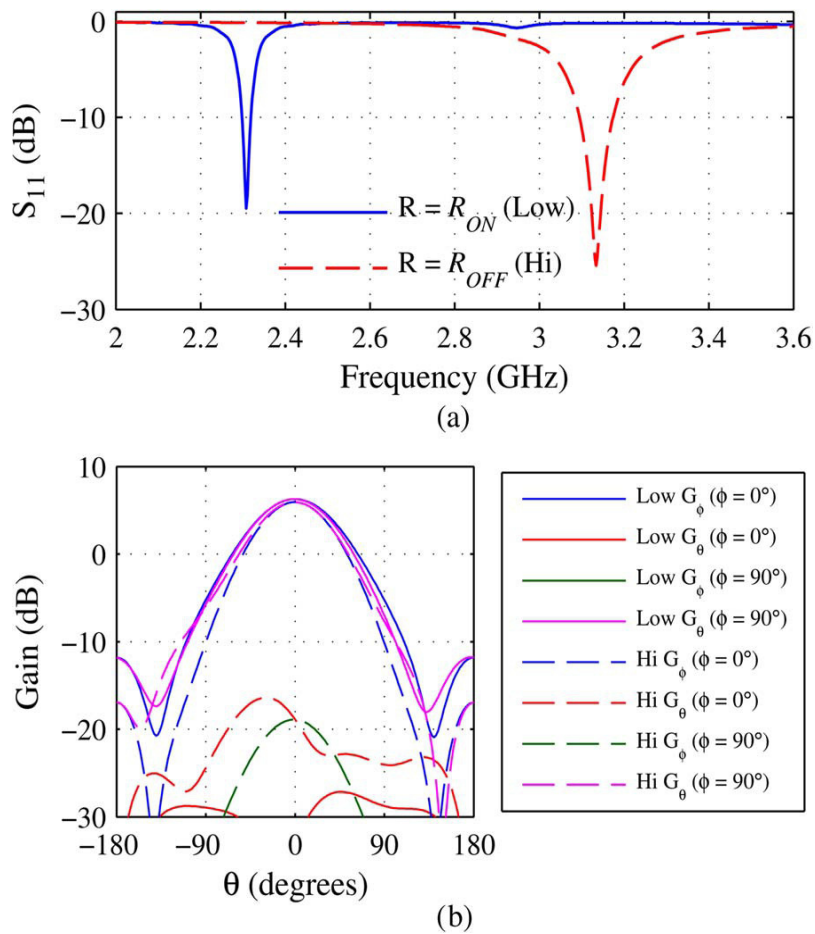

Figure 17. Scattering parameter $S_{11}$ versus frequency (a) and gain versus angles (b) for the two state of the reconfigurable antenna (*from [10]*).

At last the use of the memristor simplifies the circuit compared to those using FET as switching elements. In this case it is very important to avoid the presence of the wires needed to set the FET since these wire could either pick RF signal and propagate it to the FET gate, either interfere with the emission pattern.

## 5. CONCLUSION AND FUTURE DEVELOPMENTS

In this paper we have analyzed the mathematical properties of memristor, predicted over 40 years ago by Leon Chua. A description of the first memristor, manufactured at HP Laboratories, has been presented. HP Memristor is easy to manufacture but requires the modern nanometre scale process. Moreover we have analyzed a memristor model particularly able to be implemented in P-SPICE simulator. Finally we highlight its potential applications in the circuit design and computer technology. These applications include but are not limited to: non-volatile memory, analog electronics, image processing and reconfigurable RF antenna. These applications take advantage of the memristors ability to store digital and analog information in a simple and power efficient manner.

At HP Discover Conference in June 2014, HP announced an ambitious plan to use memristors to build a system, called simply "The Machine," shipping as soon as the end of the decade. By 2016, the company plans to have memristor-based DIMMs, which will combine the high storage densities of hard disks with the high performance of traditional DRAM.

# Authors

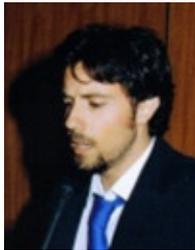

**Roberto Marani** received the Master of Science degree (*cum laude*) in Electronic Engineering in 2008 from Polytechnic University of Bari, where he received his Ph.D. degree in Electronic Engineering in 2012.

He worked in the Electronic Device Laboratory of Bari Polytechnic for the design, realization and testing of nanometrical electronic systems, quantum devices and FET on carbon nanotube. Moreover Dr. Marani worked in the field of design, modelling and experimental characterization of devices and systems for biomedical applications.

In December 2008 he received a research grant by Polytechnic University of Bari for his research activity. From February 2011 to October 2011 he went to Madrid, Spain, joining the Nanophotonics Group at Universidad Autónoma de Madrid, under the supervision of Prof. García-Vidal.

Currently he is involved in the development of novel numerical models to study the physical effects that occur in the interaction of electromagnetic waves with periodic nanostructures, both metal and dielectric. His research activities also include biosensing and photovoltaic applications.

Dr. Marani is a member of the COST Action MP0702 - Towards Functional Sub-Wavelength Photonic Structures, and is a member of the Consortium of University CNIT – Consorzio Nazionale Interuniversitario per le Telecomunicazioni.

Dr. Marani has published over 100 scientific papers.

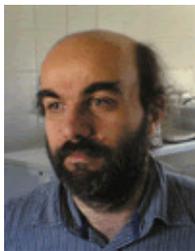

**Gennaro Gelao** received the Laurea degree in Physics from University of Bari, Italy, in 1993 and his Ph.D. degree in Physics from University of Bari in 1996, working at the ALEPH experiment, CERN.

Dr. Gelao worked at the start-up of an electrical metrological laboratory at ENEA (Rotondella, Italy)

Currently he cooperates with the Electronic Device Laboratory of Polytechnic University of Bari for the design, realization and testing of nanometrical electronic systems, quantum devices and CNTFETs. His research activity involves also the design and experimental characterization of devices and systems for biomedical applications.

Dr. Gelao has published over 80 papers.

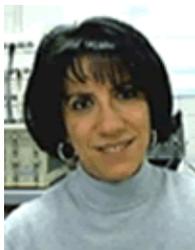

**Anna Gina Perri** received the Laurea degree *cum laude* in Electrical Engineering from the University of Bari in 1977. In the same year she joined the Electrical and Electronic Department, Polytechnic University of Bari, Italy, where she is Full Professor of Electronics from 2002.

From 2003 she has been associated with the National Institute of Nuclear Phisics (INFN) of Napoli (Italy), being a part of the TEGAF project: "Teorie Esotiche per Guidare ed Accelerare Fasci", dealing with the optimal design of resonance-accelerating cavities having very high potentials for cancer hadrontherapy.

In 2004 she was awarded the "Attestato di Merito" by ASSIPE (ASSociazione Italiana per la Progettazione Elettronica), Milano, BIAS'04, for her studies on electronic systems for domiciliary teleassistance.

Her current research activities are in the area of numerical modelling and performance simulation techniques of electronic devices for the design of GaAs Integrated Circuits and in the characterization and design of optoelectronic devices on PBG (Phothonic BandGap).

Moreover she works in the design, realization and testing of nanometrical electronic systems, quantum devices, FET on carbon nanotube and in the field of experimental characterization of electronic systems for biomedical applications.

Prof. Perri is the Head of the Electron Devices Laboratory of the Polytechnic University of Bari.

She has been listed in the following volumes: Who's Who in the World and Who's Who in Engineering, published by Marquis Publ. (U.S.A.).

She is author of over 250 journal articles, conference presentations, twelve books and currently serves as a Referee of a number of international journals.

Prof. Perri is the holder of two italian patents and the Editor of two international books.

She is also responsible for research projects, sponsored by the Italian Government.